\documentclass[]{article}
\usepackage{emulateapj}
\usepackage{graphicx}
\usepackage[usenames,dvips]{color}

\advance \voffset by -.5cm\relax

\def\msun{{\rm ~M}_{\odot}}

\def\mpy{{\rm ~M}_{\odot} {\rm ~yr}^{-1}}

\begin{document}

\title{On the Apparent Lack of Be X-ray Binaries with Black Holes}

 \author{Krzysztof Belczynski\altaffilmark{1,2,3},
         Janusz Ziolkowski\altaffilmark{4}}

 \affil{
     $^{1}$ Los Alamos National Lab,
            P.O. Box 1663, MS 466, Los Alamos, NM 87545 \\
     $^{2}$ Oppenheimer Fellow\\
     $^{3}$ Astronomical Observatory, Warsaw University, Aleje Ujazdowskie
            4, 00478 Warsaw, Poland\\
     $^{4}$ Nicolaus Copernicus Astronomical Center,
            Bartycka 18, 00-716 Warszawa, Poland;\\
     kbelczyn@nmsu.edu, jz@camk.edu.pl}

 \begin{abstract}
In the Galaxy there are $64$ Be X-ray binaries known to-date. Out of those,
$42$ host a neutron star, and for the reminder the nature of a companion
is not known. None, so far, is known to host a black hole. There seems
to be no apparent mechanism that would prevent formation or detection
of Be stars with black holes. This disparity is referred to as a missing
Be -- black hole X-ray binary problem. We point out that current
evolutionary scenarios that lead to the formation of Be X-ray binaries
predict that the ratio of these binaries with neutron stars to the
ones with black holes is rather high $F_{\rm NStoBH} \sim 10-50$, with
the more likely formation models providing the values at the high
end. The ratio is a natural outcome of {\em (i)} the stellar initial
mass function that produces more neutron stars than black holes and 
{\em (ii)} common envelope evolution (i.e.~a major mechanism involved
in the formation of interacting binaries) that naturally selects 
progenitors of Be X-ray binaries with neutron stars (binaries
with comparable mass components  have more likely survival
probabilities) over ones with black holes (which are much more likely
to be common envelope mergers). A comparison
of this ratio (i.e.~$F_{\rm NStoBH} \sim 30$) with the number of
confirmed Be -- neutron star X-ray binaries ($42$) indicates that the
expected number of Be -- black hole X-ray binaries is of the order of
only $\sim 0-2$. This is entirely consistent with the observed Galactic
sample.
 \end{abstract}

\keywords{binaries: close --- black hole physics --- stars: evolution ---
          stars: neutron}

\section{Introduction}

High mass X-ray binaries host a compact object (a neutron star or a
black hole) and a massive star (Liu, van Paradijs \& van den Heuvel
2000, 2005, 2006). The major subclass of high mass X-ray binaries
consists of a Be star and a compact object and they are referred to
as Be X-ray binaries (Be XRBs, e.g., Hayasaki \& Okazaki 2005;
2006). The Be stars are massive, generally main sequence, stars of
spectral types A0-O8 with Balmer emission lines (Zorec \& Briot 1997,
Negueruela 1998). The Be XRBs are
found with rather wide (orbital periods in the range of $\sim
10-300$ days) and frequently eccentric orbit and a compact object
accretes from the wind of a Be star (even massive Be stars are
within their Roche lobes for these wide orbits). At present, $64$ Be
XRBs are known in the Galaxy, and in $42$ the compact object was
confirmed to be a neutron star (NS) by the presence of the X-ray
pulsations (see Table~\ref{galactic}). In the remaining cases,
whenever we have information concerning the nature of the compact
component (such as an X-ray spectrum), it also indicates a NS.
Although one cannot exclude that a few of these systems contain
white dwarfs or black holes, it is fair to state that majority of
them contain NSs as compact components.

Other classes of XRBs are (with one exception) less numerous. We know
$90$ X-ray bursters (which all host NSs) and $44$ X-ray pulsars not
associated with a Be type companion ($30$ of these NSs are associated
with a supergiant type companion and $14$ with a low mass companion).
In addition, we know $57$ black hole candidate systems (among them 21
confirmed BH systems; e.g., Orosz 2003; Casares 2007; Ziolkowski 2008).
However, not a single black hole binary containing a Be type component
has been found so far. This disparity, $42$ Be XRBs with NSs versus not
a single one with a BH, seems indeed striking.

The X-ray emission from Be XRBs (with a few exceptions) is of a
distinctly transient nature with rather short active phases
separated by much longer quiescent intervals (a flaring behavior).
There are two types of flares, which are classified as Type I
outbursts (smaller and regularly repeating) and Type II outbursts
(larger and irregular; Negueruela \& Okazaki 2001, Negueruela et al. 
2001). Type I bursts are observed in systems with highly eccentric orbits. 
They occur close
to periastron passages of a NS. They are repeating at intervals
$\sim P_{\rm orb}$. Type II bursts may occur at any orbital phase.
They are correlated with the disruption of the excretion disc around
Be star (as observed in H$\alpha$ line). They repeat on time scale
of the dynamical evolution of the excretion disc ($\sim$ few to few
tens of years). This recurrence time scale is generally much longer
than the orbital period (Negueruela et al. 2001).

Be XRBs systems are known to contain two discs: excretion disc
around Be star and accretion disc around neutron star. Both discs
are temporary: excretion disc disperses and refills on time scales
$\sim$ few to few tens of years (dynamical evolution of the disc,
formerly known as the ``activity of a Be star'' (Negueruela et al. 2001), 
while the accretion disc disperses and refills on time scales $\sim$ weeks
to months (which is related to the orbital motion on an eccentric
orbit and, on some occasions, also to the major instabilities of the
excretion disc). The accretion disc might be absent over a longer period
of time ($\sim$ years), if the other disc is very weak or absent.
The X-ray emission of Be XRBs binaries is controlled by the
centrifugal gate mechanism, which, in turn, is operated both by the
periastron passages (Type I bursts) and by the dynamical evolution
of the excretion disc (both types of bursts). This mechanism
explains the transient nature of the X-ray emission (see Ziolkowski 
2002 and references therein).

One should add that the excretion discs are not a mystery any more.
In recent years, the outflowing viscous excretion discs were used to describe
the circumstellar matter around Be stars known earlier as ``an
envelope of a Be star'' (Okazaki 1997; Porter 1999; Negueruela \&
Okazaki 2001). The modeling with the help of the viscous excretion
discs appeared to be by far more successful in describing the
circumstellar matter, than earlier descriptions in terms of
``equatorial winds'', ``expanding envelopes'' or ``ejected shells''. In
particular, the viscous disc models were able to explain the very
low outflow velocities (the observed upper limits are, at most, a few
km~s$^{-1}$ and, also, to explain the (so called) V/R variability,
observed in Be stars. The viscous excretion discs are very similar
to the, well known, viscous accretion discs, except for the changed
sign of the rate of the mass flow. Some aspects of the modeling
(the supply of the matter with the sufficient angular momentum to the
inner edge of the disc, the interaction of the stellar radiation with
the matter of the disc) are not fully solved yet, but the general
picture is quite convincing. The viscosity in the excretion discs is
usually assumed (similarly as for accretion discs) to be in the form of
$\alpha$-viscosity. The discs are almost Keplerian (rotationally
supported) which explains the very low values of the radial
component of velocity. Nearly Keplerian discs (both inflowing and
outflowing) were, since a long time, known to undergo a global
one-armed oscillation instability (Kato 1983). This instability
(progressing density waves) provide a very successful explanation of
V/R variability, observed both in isolated Be stars and in members
of Be/X-ray systems. This phenomenon manifests itself in the form of
quasi-cyclical changes of the ratio of the strengths of the V(iolet)
peak to the R(ed) peak in the double profile emission lines. This
variability (best seen for the H${\alpha}$ line) includes phases,
when only one peak is visible. The time scales of the quasi-cycles
range from months to years or decades. The theoretical line profiles
calculated for the discs with an asymmetric matter distribution
(due to progressing density waves) were found to be in good
agreement with the observed profiles (Okazaki 1996; Hummel \&
Hanuschik 1997). Also the theoretical time scales calculated for
the one-armed oscillation instability agreed with the observed time
scales of V/R variability (Negueruela et al.~2001). The one-armed
instability leads ultimately to the disruption of the disc and
ejection of the matter from its outer rim. This phenomenon is
believed to be responsible for Type II bursts. Therefore, these
time scales describe also the recurrence of Type II bursts.

The total number of Be XRBs in the Galaxy is difficult to estimate.
The typical duty cycle (the relative length of the interval of the high
X-ray emission, during which the system might be detected) is rather
small: $\sim 1\%$ to $\sim 10\%$ (with the exception of a few persistent
sources). The coverage of the sky by X-ray surveys is very far from
complete. Many systems lie in the obscured regions of the Galaxy and
they might be detected only in hard X-rays. With the exception of
INTEGRAL (hard X-ray capability which has led to the discovery of
many new Be XRBs; Bird et al. 2007), these systems are undetectable by
most X-ray observatories. Since most of Be XRBs are dormant at any given
moment, we can expect to discover many new such systems in near
future. It might be estimated that the total number of Be XRBs in
the Galaxy is perhaps an order or two orders of magnitude larger
than the number of the presently known systems (e.g., Rappaport \&
van den Heuvel 1982; van den Heuvel \& Rappaport 1987).

In a series of conference proceeding papers (Sadowski et al. 2008a, 2008b,
2008c) we have reported a potential solution to the problem of the missing
black hole Be X-ray binaries. This solution was incorrect and we want to
stress to not use these papers. The errors were made in population synthesis
result analysis. In this study, we have performed the calculations anew and
we have double checked our analysis. Additionally, all the calculations were
performed with the revised code (see \S\,2.1). The new results along with
the new and qualitatively different solution to the missing problem are
presented in this study. 

In this work we study the origins of the apparent disparity of
number of known Be XRBs with NSs ($42$) as compared to no known Be
XRBs with BHs in Galaxy. This disparity has been noted in the
literature for some time. The first stellar population synthesis calculations 
intended to estimate the number of black hole Be X-ray binaries were carried 
out by Raguzova \& Lipunov (1999). They {\em assumed} that a ratio of 
Be X-ray binaries with a NS to the ones with a BH is $\sim 25$. This was based 
on $25$ Galactic NS Be X-ray binaries known at that time and the 
assumption that GRS 1915+105 is a BH Be X-ray binary. We know now that this
is not the case and that GRS 1915+105 is a low mass X-ray binary with the 
donor star being a low mass (K-M) giant (e.g., Greiner et al. 2001). 
They demonstrated that by adjusting some evolutionary parameters (e.g., natal
kicks, the initial star mass that divides a NS from a BH formation)
their model can reproduce the ratio of $\sim 25$. However, some of their 
adjustments seem to be rather extreme in the light of the current
understanding of binary evolution. For example, they needed to assume that
stars form black holes {\em only} if their mass is higher than $\sim 60
\msun$ and that {\em each} main sequence star above $10 \msun$ becomes a 
Be star after a mass accretion episode. 
Another population synthesis study connected to the Be phenomenon was
presented by Zhang et al. (2004). In this study it was noted 
that, according to the stellar population synthesis
calculations by Podsiadlowski et al.~(2003), BH binaries are 
predominantly formed with relatively short orbital periods ($P_{\rm orb} <
10$\,days). If this is the case, then, according to Zhang et al.
(2004), the excretion disc truncation mechanism (Artymowicz \& Lubow
1994) might be so efficient that the accretion rate is very low and
the system remains dormant (and therefore invisible) almost all the
time. One should note, however, that Podsiadlowski et al. (2003)
considered essentially BH systems with Roche lobe filling
secondaries, definitely not the case for Be XRBs (wind-fed
accretion; e.g. Negueruela 1998; Ziolkowski 2002).

We want to stress that in this paper we consider only Galactic Be
XRBs. However, it should be added that there exists a large
population of Be XRBs in the Magellanic Clouds (e.g., Liu et al. 2005). 
It is even larger than that of the Galaxy: $74$ known systems with  
$47$ containing a confirmed pulsar; Ziolkowski, unpublished 
compilation.
Moreover, this population is
growing very fast. Taking into account that the total stellar
content of the Magellanic Clouds is much smaller than the Galactic one,
the study of the origin of this population might be very
interesting. Due to very different properties of stellar populations
(e.g., lower metallicity, different star formation history) we defer the
study of Be XRBs in the Magellanic Clouds to another paper.

\section{Model}

\subsection{Population Synthesis}

Binary population synthesis is used to calculate the population of massive
stars (spectral types B/O) that are in binaries with a compact  object;
either a neutron star (NS) or a black hole (BH). The population synthesis
code employed in this work, {\tt StarTrack}, was initially developed
for the study of double compact object mergers in the context of
gamma-ray burst (GRB) progenitors (Belczynski, Bulik \& Rudak 2002a)
and gravitational-wave inspiral sources (Belczynski, Kalogera \& Bulik
2002b). Single stellar evolution is based mostly on the modified set of
stellar models of Hurley, Pols \& Tout (2000) while binary evolution
procedures were independently developed for this code.

In recent years {\tt StarTrack} has undergone major updates
and revisions in the physical treatment of various binary evolution
phases, and especially mass transfer phases. The new version has
already been tested and calibrated against observations and detailed
binary mass transfer calculations (Belczynski et al.\ 2008), and has
been used in various applications. The physics updates that are most
important for compact object formation and evolution include: (i) a full
numerical approach to orbital evolution due to tidal interactions,
calibrated using high mass X-ray binaries and open cluster observations;
(ii) a detailed treatment of mass transfer episodes fully calibrated against
detailed calculations with a stellar evolution code; (iii) updated stellar
winds for massive stars (see Belczynski et al.\ 2009a); and (iv)
the latest determination of the natal kick velocity
distribution for neutron stars (Hobbs et al.\ 2005). For He
star evolution, which is of crucial importance for the formation of
double neutron star binaries (e.g.~Ivanova et al.\ 2003; Dewi \& Pols
2003), we have applied a treatment matching closely the results of
detailed evolutionary calculations. If the He star expands and
Roche lobe overflow (RLOF) begins, the systems are examined for the
potential development of a dynamical instability, in which case they
are evolved through a common envelope (CE) phase, otherwise a highly
non-conservative mass transfer ensues. We treat CE events using the
energy formalism (Webbink 1984), where the binding energy of the
envelope is determined from the set of He star models calculated
with the detailed evolutionary code by Ivanova et al.\ (2003). The
progenitor evolution and Roche lobe overflow episodes are now
followed in much greater detail. In particular the code was calibrated
for various cases of mass transfer in massive binaries and then tested
against some published detailed evolutionary tracks (e.g.~Wellstein,
Langer \& Brown 2001).

One specific evolutionary process that plays an important role for
the formation of binaries with black holes is the CE phase.
This phase is instrumental in decreasing orbital separation and bringing
initially distant binary components close to each other. This in turn
allows components of many systems to begin an interaction and manifest their
presence through a number of observational phenomena.
The inspiral process in common envelope and the subsequent decrease of
an orbit is not well understood. It is believed that only donor stars
(initiating the process) with the well developed core-envelope
structure, like red giants, can successfully survive the inspiral. On
the other hand stars with no clear core-envelope boundary, like main
sequence stars, can not survive this phase, leading to a CE
merger and formation of one, rapidly rotating and peculiar
star.

It was pointed out that stars crossing the Hertzsprung gap (HG)
do not have a clear entropy jump at the core-envelope transition
(Ivanova \& Taam 2004). Therefore, it can be expected that if such a
star overflows its Roche lobe and initiates a CE phase the inspiral will
eventually lead to a merger (e.g. Taam \& Sandquist 2000).
This possibility was tested in evolutionary calculations of merger
rates of double compact objects. It turns out that if such a possibility
is accounted for, the merger rates of BH-BH binaries decrease drastically
(by factor of $\sim 500$). NS-NS merger rates drop only
moderately (Belczynski et al. 2007). Since the details of the CE
phase are not fully understood, in this study we will also test this
possibility in our predictions for Be X-ray binaries. In model A,  
we will allow for survival even if a donor star is on the HG (i.e.,
standard energy balance is calculated to check on the outcome of CE),
while in model B, any CE phase that involves an HG donor is {\em
assumed} to lead to a merger aborting subsequent binary evolution and
potential formation of a Be X-ray binary.

Another important factor that determines the formation efficiencies of
binaries with neutron stars and black holes is the natal kick compact
objects receive at birth. For models A and B we adopt natal kicks from
the radio pulsar birth velocity distribution derived by Hobbs et al.
(2005; a Maxwellian with $\sigma=265$~km~s$^{-1}$) for neutron stars
that are formed in regular core-collapse supernovae, while for neutron
stars formed in electron capture supernovae we adopt no natal kicks
(e.g.~Dessart et al.~2006). Natal kicks for black holes are decreased
proportionally to the amount of fall back expected during core
collapse/supernova explosion (e.g.~Fryer 1999; Fryer \& Kalogera
2001). However, we calculate one extra model in which natal kicks are
reduced. In particular, in model C, neutron star kicks are drawn
from a Maxwellian distribution with $\sigma=133$~km~s$^{-1}$, while black hole
kicks are derived from the same distribution but then decreased due to
the fall back. For the electron capture neutron star formation, no
natal kicks are applied. There are some indications that the natal kicks
neutron stars receive are smaller for stars in binaries as compared with 
single stars (e.g.~Podsiadlowski et al.~2004). Here we have
decreased the value of kicks as measured by Hobbs et al.~(2005) for
single pulsars by half and applied them to neutron stars in
binaries. This specific choice was motivated by the recent calculation
that demonstrated that in order to reproduce the ratio of recycled
pulsars in double neutron star binaries (9 known) to single recycled
pulsars (4) the kicks operating in binary stars are required to be
rather low ($\sigma \lesssim 133$ km s$^{-1}$; Belczynski et
al. 2009b).

Delineation in the formation of neutron stars and black holes may  
play an important role in the expected number of Be X-ray binaries with 
different accretors. In particular, we still do not know what a star in 
the initial mass range of $M_{\rm zams}=20-40 \msun$ forms: a neutron 
star or a black hole? In this study we have employed a physical model 
for the formation of compact object, and rather than using a star initial
mass (as most population synthesis codes do), we have used the star 
properties to delineate the formation of a neutron star from the formation 
of a black hole. In short, we follow the evolution of a given star, note 
the final mass of its FeNi core and then we use the results of hydro 
core-collapse simulations to estimate the mass of the compact object a 
given star forms. Once we have the mass of a compact object we use the 
maximum neutron star mass (assumed to be $2.5 \msun$) to tell apart 
neutron stars from black holes. The full description of this scheme is 
given in Belczynski et al. (2008; and see the references within). As it 
happens our scheme (for single stars) results in the black hole formation 
for $M_{\rm zams} \gtrsim 20 \msun$ (see Belczynski et al. 2009b). If we 
wanted to (artificially) adopt a higher limit (e.g., $M_{\rm zams} 
\gtrsim 40 \msun$), we would expect more Be X-ray binaries with neutron
stars than predicted in the following sections. In other words, our
conclusions are rather conservative on the estimate of the ratio of 
Be X-ray binaries with neutron stars to the ones with black holes. If
anything such a ratio should be higher and our conclusions that follow
stronger.

\subsection{Calculations}

We evolve a Galactic population of massive binaries using {\tt StarTrack}.
We adopt solar metallicity ($Z=0.02$), and a steep initial mass function
(IMF) for massive stars with a power-law exponent of $-2.7$ (Kroupa, Tout
\& Gilmore 1993; Kroupa \& Weidner 2003). Roche lobe overflow is treated
in a non-conservative way (with $50\%$ mass loss from a given binary;
e.g.~Meurs \& van den Heuvel 1989) while the CE phase is treated via energy
balance with fully efficient transfer of orbital energy into dispersal of
an envelope (e.g.~$\alpha \times \lambda =1.0$).
The results are calibrated in such a way that the Galactic star
formation rate is at the level of $3.5 \mpy$ and is constant through the last
10~Gyr (e.g.~O'Shaughnessy et al. 2006). At the present Galactic disk age
($t=10$ Gyr) we perform a time slice and extract Be X-ray binaries using
classification criteria defined in the following section.

\subsection{Be X-ray Binary}

Following the earlier discussion (\S\,1) of the observed properties of
Galactic population of Be X-ray binaries, we adopt a definition (that is
extended to potential systems that may host a BH) of a Be X-ray binary
in our population synthesis calculations. We call any system a Be X-ray
binary if:
{\em (i)} it hosts either a NS or a BH accretor;
{\em (ii)} donor is a main sequence star (burning H in its core);
{\em (iii)} donor mass is high $M_{\rm don} \geq 3.0 \msun$ (O/B star);
{\em (iv)} accretion proceeds {\em only} via stellar wind (no RLOF);
{\em (v)} orbital period is in the range $10 \leq P_{\rm orb} \leq 300$ day; and
{\em (vi)} only a fraction $F_{\rm Be}=0.25$ of the above systems are designated
as hosting a Be star and not a regular O/B star. The observations indicate
that the fraction of Be stars among all B stars is $1/5$ to $1/3$ (e.g.,
Slettebak 1988; Ziolkowski 2002; McSwain \& Gies 2005).
Note that in the above definition we do not require a non-zero eccentricity
as for the Galactic BeXRBs as many as $\sim 20\%$ systems have small
eccentricity ($\lesssim 0.1$).

We have chosen to define Be X-ray binary in a phenomenological way, and
based on the observational properties we have adopted the values for our
limiting factors (e.g., $F_{\rm Be}$). Note also, that we do not have put any
constrains on the spin of a Be star. It translates into an assumption (as Be
stars are known to have high rotational velocities; e.g., Slettebak 1949, 1966;
Slettebak, Collins \& Truax 1992) that either some stars are born with initially
high spins or that the mass accretion during RLOF can effectively spin up a
massive star. The more physical approach (e.g., B star becomes Be star if spun
up via RLOF accretion) is also potentially possible within the framework of our
population synthesis model.
However, the details of such a model would be highly uncertain (e.g., the high
spin of Be stars may be connected to their initial rapid rotation and not to
RLOF spin up; it is not certain if any spin up is expected during CE phase; and
how effectively angular momentum can be transferred during RLOF).
We note that the adopted value of $F_{\rm Be}$ affects only the absolute
number of Be X-ray binaries predicted in our calculations. However, the
physical properties or the relative ratio of Be X-ray binaries hosting a NS
to the ones hosting a BH remains the same and therefore our results and
conclusions (see below) are independent of the value we adopt for $F_{\rm Be}$.

\section{Results}

\subsection{Models A and B}

In Table~\ref{channels}, the main results of our calculations are presented.
The main formation channels for Be X-ray binaries are given separately
for systems with NSs (marked $BeNS:0N$ where $N$ is a number that indicates
a given channel) and BHs ($BeBH:0N$). We also list the predicted intrinsic
Galactic number of Be X-ray binaries ($N_{\rm BeNS},\ N_{\rm BeBH}$), along
with the expected ratio of systems with NSs to systems with BHs
($F_{\rm NStoBH}$).

Binaries with NS form mostly along two channels. One ($BeNS:01$; that forms
$\sim 45\%$ of Be X-ray binaries with NSs) involves a CE phase when the
primary is
an evolved star (e.g.~burning He in its core, or burning H and He in a shell)
and after losing the
envelope, the combination of the post-CE orbital separation
(rather large) and maximum radius of exposed primary core (not too large) is
such that there are no more interactions until the primary's exposed core
explodes and forms a NS. The other channel ($BeNS:02$; $\sim 45\%$) involves
also the initial CE phase, but it usually starts on a smaller orbit, so the
post-CE separation is such that it allows the exposed core of primary (naked
He star) to overflow its Roche lobe and start another interaction (non
conservative mass transfer) before it explodes and forms a NS.
There is also one channel ($BeNS:03$; $\sim 10\%$) that involves two binary
components of comparable mass, and once RLOF begins it does not develop
into a CE phase but proceeds via regular and not too violent way (on a
thermal and/or nuclear timescale of the donor). After the RLOF episode the
primary explodes and forms a NS.
At the time of NS formation the companion is a still unevolved (main sequence)
star with high mass ($M_{\rm b}  \geq 3.0 \msun$). There is not much
difference from model A to B in the formation of Be X-ray binaries with NSs.

The formation of Be X-ray binaries with BHs proceeds along two channels.
The first one ($BeBH:01$) is very similar to $BeNS:01$ and
involves a CE phase followed by the explosion of a primary and the formation
of a BH. The main difference comes from the fact that in this case the
primary is more massive (than a progenitor of a NS in $BeNS:01$) and it
starts a CE phase earlier in its evolution. The CE phase is encountered very
often when a primary is crossing the HG, as it is the first
evolutionary phase at which the stars undergo a very significant radial
expansion. Since we impose non survival in such a case for our model B, we
note the significant decrease of formation of Be X-ray binaries in this
channel from model A ($\sim 80\%$) to model B ($\sim 10\%$).
The second formation channel ($BeBH:02$) of Be X-ray binaries with BHs is
almost the same as the channel $BeNS:03$ for binaries with NS. Since this
channel involves only one non-conservative RLOF, the change from model A
($\sim 20\%$) to model B ($\sim 90\%$) is only relative. This change simply
reflects the decrease of number of binaries in channel $BeBH:01$ and the
actual number of binaries in the channel $BeBH:02$ does not change from
model A to B.

We also present the predicted numbers of Be X-ray binaries calibrated for
the Galactic disk star formation rate. We have selected (randomly) only
$25\%$ ($F_{\rm Be}=0.25$) of the massive binaries hosting B/O star with a
compact object companion as this seems to be the fraction of Be stars among
regular B stars and we have additionally imposed some constraints on orbital
period and system configuration (see \S\,2.3). The number of predicted Be X-ray
binaries with NSs is
$N_{\rm BeNS} \sim 500$ and does not depend much on the adopted model of CE
evolution. The number of Be X-ray binaries with BHs is significantly smaller
and is sensitive to the adopted CE model; we find $N_{\rm BeBH} \sim 80$ and
$\sim 20$ systems for model A and model B, respectively. This is
qualitatively the same trend as found by Belczynski et al.~(2007), who
demonstrated that CE survival/non-survival affects mostly systems with BHs.

In each model we predict more Be XRBs with NSs than with BHs.
In general there are more NSs than BHs ($\sim 8:2$; the ratio that depends on
adopted IMF and the details of the transition from NS to BH formation).
Additionally, binaries with initially comparable mass components (like
progenitors of Be X-ray binaries with NSs) are more likely to survive
the first interaction (usually CE phase) than the binaries with
components of very different masses (like for many progenitors of Be
X-ray binaries with BHs).  The number of Be XRBs with NSs is about the
same in both models, but the number of binaries with BHs decreases
significantly from model A to B. Therefore, the ratio of Be XRBs with
NSs to the ones with BHs changes significantly from model A ($F_{\rm
NStoBH}=7$) to model B ($F_{\rm NStoBH}=27$).

\begin{figure}
\includegraphics[width=1.0\columnwidth]{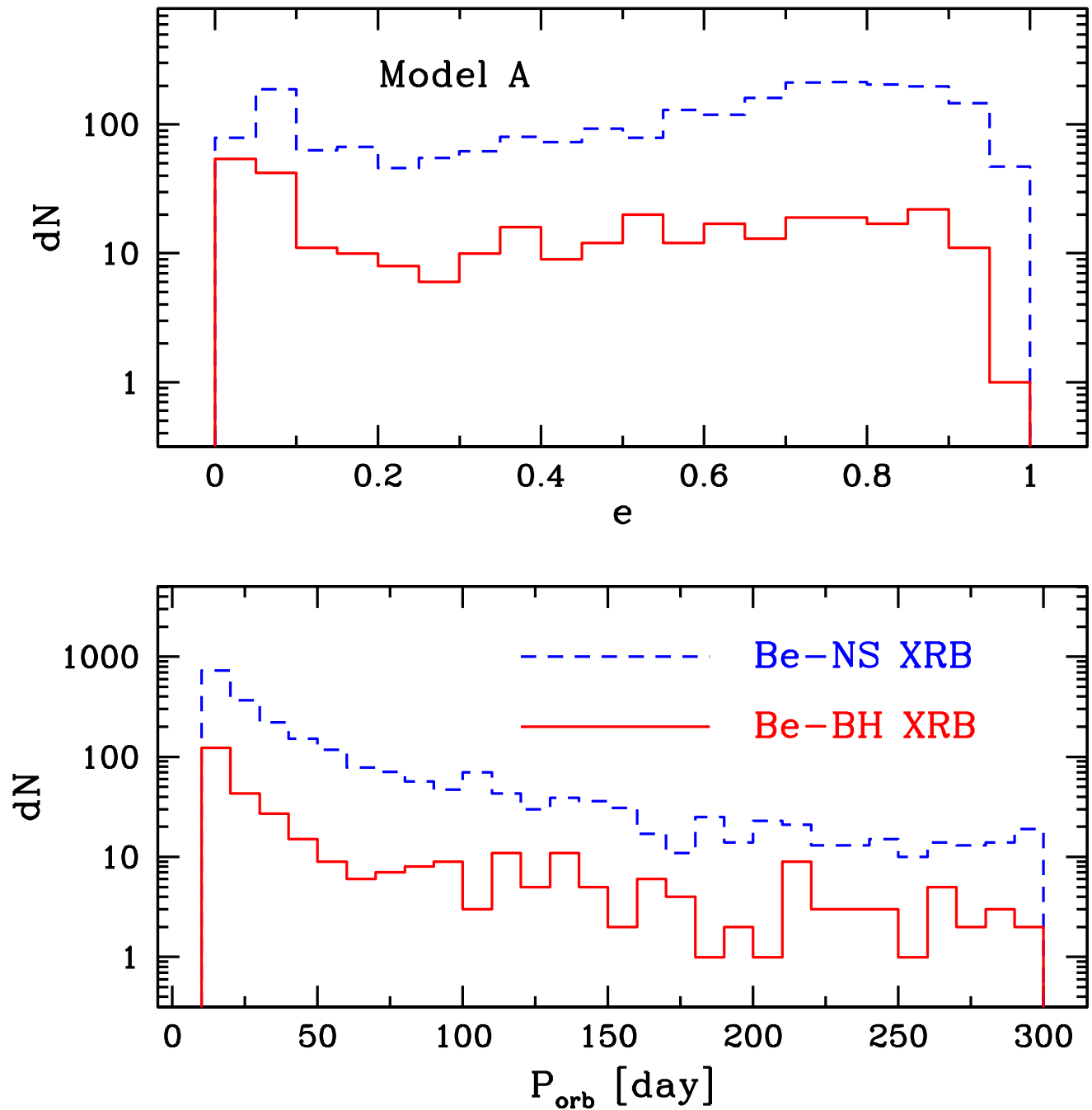}
\caption{Orbital period and eccentricity distributions for different
populations of Be X-ray binaries: systems with neutron stars are more
abundant that the ones with black hole accretors. However, the distribution
of periods is rather similar for both populations. Distributions correspond
to model A in which survival through a common envelope phase with a
Hertzsprung gap donor is allowed (for details see \S\,2.1). Note that for
the sake of presentation we show the entire population (i.e.,
$F_{\rm Be}=1.0$) of potential Be X-ray binaries, and the actual synthetic
Galactic population as presented in Table~\ref{channels} was obtained by
drawing only $1/4$ of systems from these distributions
(i.e.~$F_{\rm Be}=0.25$).}
\label{periodA}
\end{figure}

\begin{figure}
\includegraphics[width=1.0\columnwidth]{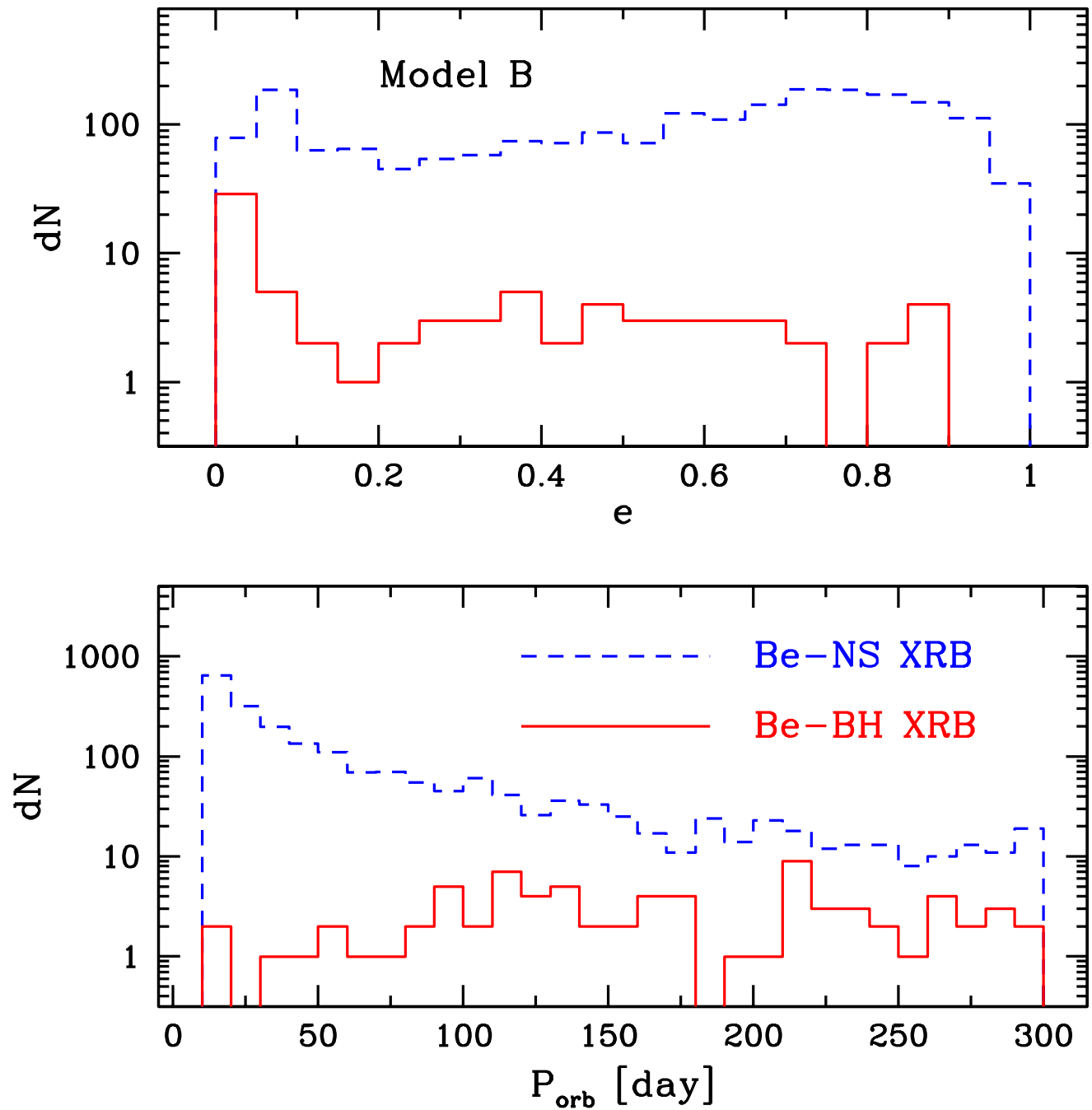}
\caption{Same as Fig.~\ref{periodA} but for model B in which survival
through a CE phase with a HG donor is not allowed. Note the significantly
smaller number of Be-BH X-ray binaries, and moderately smaller number of
Be-NS X-ray binaries.
}
\label{periodB}
\end{figure}

In Figs.~\ref{periodA} and ~\ref{periodB} we present the distributions of
orbital periods and eccentricities for the synthetic Be X-ray binaries for
models A and B. Due to the small number of binaries predicted in the
Galaxy, the distributions are constructed from the entire
population of potential Be X-ray binaries (i.e.~$F_{\rm Be} = 1.0$). The
actual population of synthetic Galactic Be X-ray binaries as listed in
Table~\ref{channels} (i.e.~$F_{\rm Be}=0.25$) is constructed by random
drawing from the distributions shown in Figs.~\ref{periodA} and
~\ref{periodB}.

Orbital periods are contained within $10 \leq P_{\rm orb} \leq 300$ days by
our adopted definition of Be X-ray binaries (see \S\,2.3). The distributions
within these limits are falling off with increasing periods and this general
trend holds for both Be X-ray binaries with NSs and BHs. The relatively
large numbers of systems with small periods are the outcome of the orbital
contraction at CE phase. This is indirectly, but very clearly, demonstrated
on the example of Be X-ray binaries with BHs in Model B (Fig.~\ref{periodB})
in which the majority systems do not evolve through a CE phase and there are
almost no binaries with small periods (e.g., with $P_{\rm orb} \leq 50$ days).

The eccentricity distributions are similar for models A and B. For Be X-ray
binaries with NSs there is an overall trend of distributions with rising
eccentricity. This is an effect of natal kicks that are rather high
($\sigma = 265$ km s$^{-1}$) so binaries in general receive high kicks and
gain significant eccentricity after the first supernova. It is noted that
we only plot here the tail of entire population of initial binaries that
may have became Be X-ray binaries, but were disrupted in the first supernova.
Actually, the disruption at the first supernova for massive binaries is very
high ($\gtrsim 95\%$) and only a very few binaries survive.
There is also an accumulation of systems with very small eccentricities
($e \lesssim 0.1$) and this is due to the population of neutron stars that
form through electron capture supernovae for which we have assumed no natal
kicks, so the non-zero (but small) eccentricities for these systems arise
from mass loss only.

The general trends are very similar for distributions of eccentricities for
Be X-ray binaries with BHs. However, we note two natural changes in relative
strengths of the trends as compared to distributions of binaries with NSs.
First, the increase of distribution with eccentricity is smaller for binaries
with BHs, as these receive, on average, lower natal kicks than NSs (due to
the fall back and smaller explosion energies in BH formation). Second, the
contribution of BH binaries with small eccentricities is higher than for
binaries with NSs. The most massive black holes form through direct collapse
(or almost full fall back) for which we assume no (or rather small) natal
kicks and since there is no (or almost no) mass loss, these systems end up
with zero (or very small) eccentricities.

\subsection{Model C}

\begin{figure}
\includegraphics[width=1.0\columnwidth]{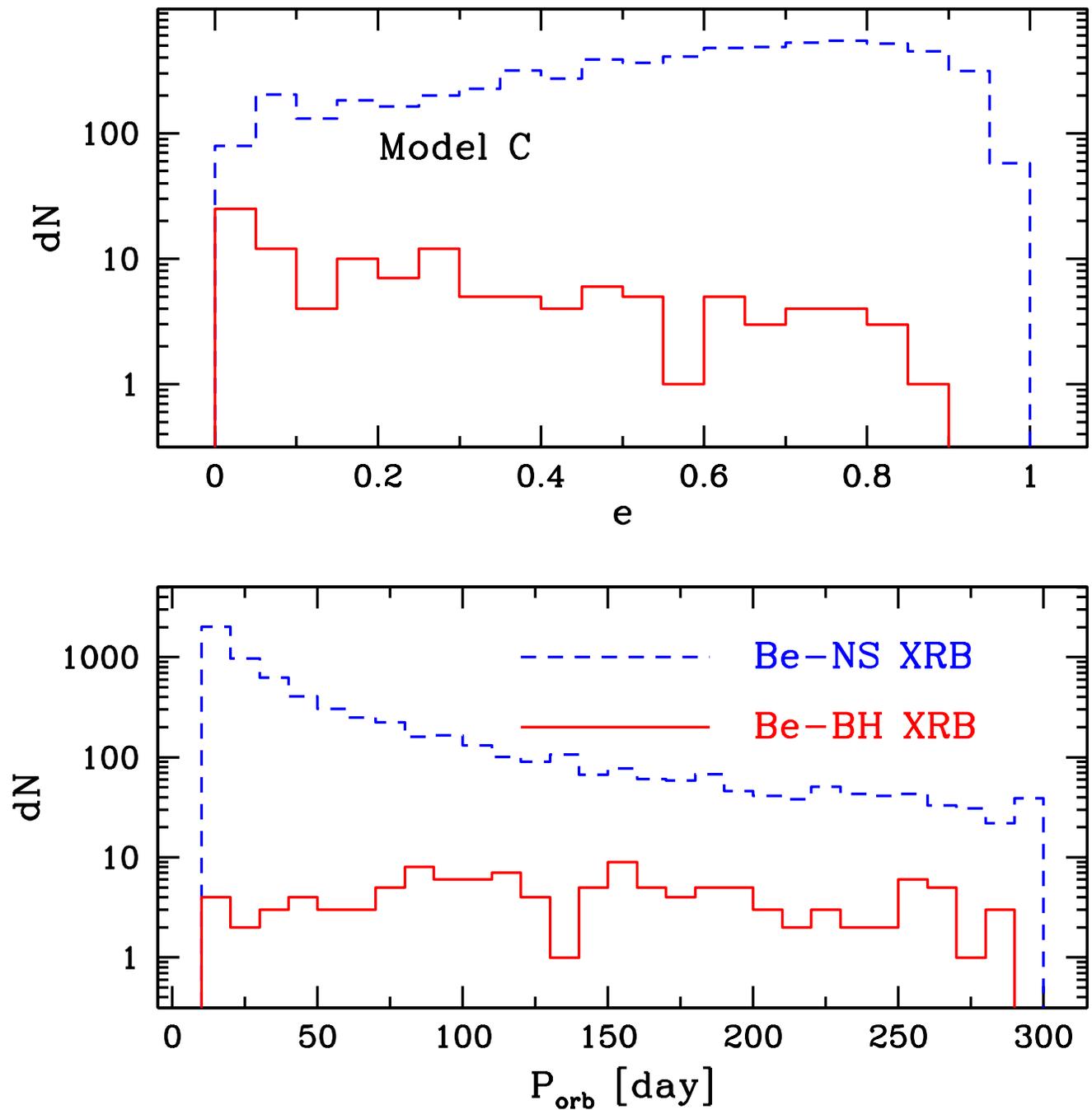}
\caption{Same as Fig.~\ref{periodA} but for model C in which the natal
kicks are decreased by half as compared with models A and B. In this model,
the survival through a CE phase with a HG donor is not allowed. Note the
significantly larger number of Be-NS X-ray binaries, and the moderately
larger number of Be-BH X-ray binaries.
}
\label{periodC}
\end{figure}

In model C we have decreased the natal kicks NSs and BHs receive and this
model is similar to model B as in this calculation we do not allow for the
CE survival if a donor is a HG star. The formation channels and corresponding
efficiencies are about the same as for model B.
However, as intuitively expected, the formation of binaries with NSs increases
($N_{\rm BeNS} \sim 1500$) significantly as compared to other models (see
Table~\ref{channels}). Since most progenitors of Be XRBs are disrupted at the
formation of a compact object (via supernova natal kick) once the kicks are
decreased the formation is more efficient. The same effect is found for Be
XRBs with BHs ($N_{\rm BeNS} \sim 30$), although at a lesser degree as BH
kicks are assumed to be smaller and therefore the disruptions are not as
important a factor as for binaries with NSs.
This trend significantly increases number of Be XRBs with NSs and only
moderately increases the number of the binaries with BHs. It therefore
has a deep impact on
the expected ratio of one population to the other. In particular, the ratio
of Be XRBs with NS to ones with BHs is found very high for model C:
$F_{\rm NStoBH}=54$.
In Fig.~\ref{periodC} we show period and eccentricity distributions for
model C.

\section{Conclusions}

We have performed a population synthesis study of Galactic Be X-ray
binaries. In particular, we have attempted to explain the problem of the
missing Be X-ray binaries with BHs. The only known Be X-ray binaries with a
confirmed type of compact object host NSs. Specifically, we know 64 Be X-ray
binaries in the Galaxy, but only 42 of these systems are known to host a NS.
None of the observed Be X-ray binaries hosts a BH.

Our main results may be described as follows:
\begin{itemize}
\item Previously, we have reported (Sadowski et al. 2008a, 2008b, 2008c)
      the potential solution to the problem of the missing Be X-ray
      binaries. This solution was incorrect and the mistake was due to
      some errors in population synthesis data analysis. The new results
      (double-checked and obtained with the updated code) are presented
      below.
\item We predict that both population of Be X-ray binaries should exist in
      the Galaxy: those with NSs as well as these with BHs.
\item The predicted number of Be X-ray binaries with NSs is much higher
      (factors of $F_{\rm NStoBH} \sim 10-50$) than those with BHs.
\item If we use the preferred evolutionary models ($F_{\rm NStoBH} \sim
      30-50$; models B and C) we predict that in the observed sample of Be X-ray
      binaries of 42 systems with NSs, one should expect only $\sim 0-2$ systems
      with BHs. It is quite possible that none are yet observed (small
      statistics).
\item Due to a very low number of expected binaries with BHs, it is very likely
      that there is no problem with missing Be X-ray binaries with BHs in Galaxy.
\end{itemize}

\acknowledgements We would like to thank anonymous referee, Duncan
Lorimer, Thomas Maccarone and Andrzej Zdziarski for a number of
useful comments on this study. Additionally, KB is indebted to
Karolina Tarczynska for her unceasing physical rehabilitation
support that made the work over this study possible.
JZ acknowledges the support from the Polish Ministry of Science and 
Higher Education (MSHE) project 362/1/N-INTEGRAL (2009-2012) and KB 
acknowledges the support from MSHE grant N N203 302835 (2008-2011).

\clearpage

\begin{deluxetable}{crrcrlr}
\tablewidth{500pt} \tablecaption{Galactic Be X-ray Binaries\tablenotemark{a}}
\tablehead{
     & $P_{\rm orb}$ & $P_{\rm spin}$ &    e & $L_{\rm x,max}$\tablenotemark{b} & Spectral & \\
Name & [d]           & [s]            &      & [erg/s]
& type     & Ref\tablenotemark{c} }

\startdata

2S 0053+604  & 203.59 &                & 0.26 & $3.9\times10^{34}$    & B0.5 Ve & 1,2 \\

4U 0115+634  & 24.3 & 3.61               & 0.34 & $3.0\times10^{37}$    & B0.2 Ve & 1,2 \\

IGR J01363+6610  &  &                &  & $1.3\times10^{35}$    & B1 Ve & 1,2 \\

RX J0146.9+6121  &  & 1404.2               &  & $3.5\times10^{35}$    & B1 Ve & 1,2 \\

IGR J01583+6713  &  &                &  &     & Be & 1 \\

1E 0236.6+6100  & 26.496 &                & 0.55 & $2.0\times10^{34}$    & B0 Ve & 1,2,3 \\

V 0332+53  & 34.25 & 4.4               & 0.37 & $> 1.0\times10^{38}$    & O8.5 Ve & 1,2 \\

4U J0352+309  & 250.3 & 837.0               & 0.11 & $3.0\times10^{35}$    & O9.5 IIIe - B0 Ve & 1,2 \\

RX J0440.9+4431  &  & 202.5               &  & $3.0\times10^{34}$    & B0.2 Ve & 1,2 \\

EXO 051910+3737.7  &  &                &  & $1.3\times10^{35}$    & B0 IVpe & 1,2 \\

1A J0535+262  & 111.0 & 103.4               & 0.47 & $2.0\times10^{37}$    & O9.7 IIIe & 1,2,4 \\

1H 0556+286  &  &                &  &     & B5ne & 1 \\

IGR J06074+2205  &  &                &  &     & B0.5 Ve & 1,5 \\

SAX J0635.2+0533  & 11.2 & 0.0338               &  & $9\div35\times10^{33}$    & B1 IIIe - B2 Ve & 1,2 \\

XTE J0658-073  &  & 160.4               &  & $6.6\times10^{36}$    & O9.7 Ve & 1,2 \\

3A J0726-260  & 34.5 & 103.2               &  & $2.8\times10^{35}$    & O8-9 Ve & 1,2 \\

1H 0739-529  &  &                &  &     & B7 IV-Ve & 1 \\

1H 0749-600  &  &                &  &     & B8 IIIe & 1 \\

RX J0812.4-3114  & 81.3 & 31.8851               &  & $1.1\times10^{36}$    & B0.2 IVe & 1,2 \\

GS 0834-430  & 105.8 & 12.3               & 0.12 & $1.1\times10^{37}$    & B0-2 III-Ve & 1,2 \\

GRO J1008-57  & 247.5 & 93.5               & 0.66 & $2.9\times10^{35}$    & B0e & 1,2 \\

RX J1037.5-5647  &  & 862.0               &  & $4.5\times10^{35}$    & B0 III-Ve & 1,2 \\

1A 1118-615  &  & 407.68               &  & $5.0\times10^{36}$    & O9.5 Ve & 1,2,6 \\

IGR J11435-6109  & 52.46 & 161.76               &  &     & Be & 1,7 \\

2S 1145-619  & 187.5 & 292.4               & $>$ 0.5 & $7.4\times10^{34}$    & B0.2 IIIe & 1,2 \\

1H 1253-761  &  &                &  &     & B7 Vne & 1 \\

1H 1255-567  &  &                &  &     & B5 Ve & 1 \\

4U 1258-61  & 132.5 & 272.0               & $>$ 0.5 & $1.0\times10^{36}$    & B0.7 Ve & 1,2 \\

2RXP J130159.6-635806  &  & 704.0               &  & $5.0\times10^{35}$    & Be ? & 1,2 \\

SAX J1324.4-6200  &  & 170.84               &  &     & Be ? & 1 \\

1WGA J1346.5-6255  &  &                &  & $6.6\times10^{32}$    & B0.5 Ve & 1,2 \\

2S 1417-624  & 42.12 & 17.6               & 0.446 & $8.0\times10^{36}$    & B1 Ve & 1,2 \\

SAX J1452.8-5949  &  & 437.4               &  & $8.7\times10^{33}$    & Be ? & 1,2 \\

XTE J1543-568  & 75.56 & 27.12               & $<$ 0.03 & $> 1.0\times10^{37}$    & Be ? & 1,2 \\

2S 1553-542  & 30.6 & 9.26               & $<$ 0.03 & $7.0\times10^{36}$    & Be ? & 1,2 \\

IGR J15539-6142  &  &                &  & $3.3\times10^{33}$    & B2-3 Vne & 1,2 \\

IGR J16207-5129  &  &                &  & $1.3\times10^{34}$    & B8 IIIe & 1,2 \\

SWIFT J1626.6-5156  &  & 15.37               &  &     & Be  & 2 \\

AX J170006-4157  &  & 714.5               &  & $7.2\times10^{34}$    & Be ? & 1,2,8 \\

RX J1739.4-2942  &  &                &  &     & Be ? & 1 \\

RX J1744.7-2713  &  &                &  & $1.8\times10^{32}$    & B0.5 V-IIIe  & 1,2 \\

\enddata
\tablenotetext{a}{Note that we have not included the following systems:
 1H 1249-637 (compact companion is probably a white dwarf, Liu et al. 2006); 
 IGR J16318-4848 (optical component is B[e] with strong supergiant wind
 and not a typical Be star, Liu et al. 2006);
 1H 1555-552 (Herbig Ae/Be optical component, Liu et al. 2006); 
PSR B1259 (non-accreting XRB, Tavani \& Arons 1997)} 
\tablenotetext{b}{maximum X-ray luminosity}
\tablenotetext{c}{ (1) Liu et al. 2006; (2) Raguzova 2007; (3) Grundstrom et al. 2007; 
(4) Smith et al. 2005; (5) Reig \& Zezas 2009 (6) Mangano 2009; (7)
Tomsick et al. 2007; (8) Torii et al. 1999; (9) Markwardt et al.
2008; (10) Beklen \& Finger 2009.} 
\label{galactic}
\end{deluxetable}
\clearpage

\setcounter{table}{0}
\begin{deluxetable}{crrcrlr}
\tablewidth{500pt} \tablecaption{Galactic Be X-ray Binaries (cont.)}
\tablehead{
     & $P_{\rm orb}$ & $P_{\rm spin}$ &    e & $L_{\rm x}^{\rm max}$ & Spectral & \\
Name & [d]           & [s]            &      & [erg/s] & type     & Ref}

\startdata

AX J1749.2-2725  &  & 220.38               &  & $2.6\times10^{35}$    & Be ? & 1,2 \\

GRO J1750-27  & 29.8 & 4.45               &  &     & Be ? & 1 \\

1XMM J180816.8-191940  &  &                &  & $1.3\times10^{33}$    & Be ? & 2 \\

AX J1820.5-1434  &  & 152.26               &  & $9.0\times10^{34}$    & O9.5 - B0 Ve & 1,2 \\

XTE J1824-141  &  & 120.0               &  &     & Be ? & 9 \\

1XMM J183327.7-103523  &  &                &  & $1.6\div7.5\times10^{32}$    & B0.5 Ve & 2 \\

1XMM J183328.7-102409  &  &                &  & $3.3\times10^{32}$    & B1-1.5 IIIe & 2 \\

GS J1843+00  &  & 29.5               &  & $3.0\times10^{37}$    & B0-2 IV-Ve & 1,2 \\

2S 1845-024  & 242.18 & 94.8               & 0.88 & $6.0\times10^{36}$    & Be ? & 1,2 \\

XTE J1858+034  &  & 221.0               &  &     & Be ? & 1 \\

4U 1901+03  & 22.58 & 2.763               & 0.036 & $1.1\times10^{38}$    & Be ? & 1,2 \\

XTE J1906+09  & 28.0 ? & 89.17               &  &     & Be ? & 1 \\

1H 1936+541  &  &                &  &     & Be & 1 \\

XTE J1946+274  & 169.2 & 15.8               & 0.33 & $5.4\times10^{36}$    & B0-1 IV-Ve & 1,2 \\

KS 1947+300  & 40.415 & 18.76               & 0.03 & $2.1\times10^{37}$    & B0 Ve & 1,2 \\

W63 X-1  &  & 36.0               &  &     & Be ? & 1 \\

EXO 2030+375  & 46.02 & 41.8               & 0.41 & $1.0\times10^{38}$    & B0 Ve & 1,2 \\

RX J2030.5+4751  &  &                &  & $1.7\times10^{33}$    & B0.5 V-IIIe  & 1,2 \\

GRO J2058+42  & 55.03 & 198.0               &  & $2.0\times10^{36}$    & O9.5-B0 IV-Ve & 1,2 \\

SAX J2103.5+4545  & 12.68 & 358.61               & $\sim$ 0.4 & $3.0\times10^{36}$    & B0 Ve & 1,2 \\

1H 2138+579  &  & 66.33               &  & $9.1\times10^{35}$    & B0-2 IV-Ve & 1,2,10 \\

1H 2202+501  &  &                &  &     & Be & 1 \\

SAX J2239.3+6116  & 262.6 & 1247.0               &  & $\sim 2.3\times10^{36}$    & B0 V - B2 IIIe & 1,2 \\
\enddata

\label{galactic}
\end{deluxetable}
\clearpage

\begin{deluxetable}{lrl}
\tablewidth{350pt}
\tablecaption{Simulations: Be X-ray Binary Formation Channels}
\tablehead{ Formation & Efficiency \tablenotemark{a} &  \\
            Channel   & Model A (B) [C]            & Evolutionary History \tablenotemark{b}}
\startdata

BeNS:01 & 44.2 (41.8)\ [45.3] \%& CE:a$\rightarrow$b, SN:a \\
BeNS:02 & 42.3 (43.9)\ [45.0] \%& CE:a$\rightarrow$b, NC:a$\rightarrow$b, SN:a \\
BeNS:03 & 11.9 (13.3)\ [8.8] \%& NC:a$\rightarrow$b, SN:a \\
BeNS:04 &  1.6\ \ \ (1.0)\ [0.9] \%& all other \\
&&\\

BeBH:01 & 79.6 (13.2)\ [17.2]  \%& CE:a$\rightarrow$b, SN:a \\
BeBH:02 & 19.8 (85.5)\ [82.8] \%& NC:a$\rightarrow$b, SN:a \\
BeBH:03 &  0.6\ \ \ (1.3)\ [0.0]\%& all other \\
&&\\

N$_{\rm BeNS}$   & 579 (517) \ [1578] & Galactic number of NS Be XRBs \\
N$_{\rm BeBH}$   &  82\ \  (19)\ [29] & Galactic number of BH Be XRBs \\
F$_{\rm NStoBH}$ &   7\ \ (27)\ [54]  & number ratio of NS to BH Be XRBs \\

\enddata
\label{channels}
\tablenotetext{a}{Efficiency for models with standard kicks
($\sigma=265$ km s$^{-1}$) in which survival through a CE
phase with a HG donor is allowed (A) and not allowed (B).
Model C shows results for evolution with small kicks ($\sigma=133$ km
s$^{-1}$) and the survival in CE with HG donors is not allowed.}
\tablenotetext{b}{Sequences of different evolutionary phases for the primary
(a) and the secondary (b):
non-conservative mass transfer (NC), CE, and
supernova explosion/core collapse event (SN).
Arrows mark direction of mass transfer episodes.}
\end{deluxetable}

\clearpage

\clearpage

\clearpage

\end{document}